\documentclass[onecolumn,floatfix,prd,aps,10pt]{revtex4-2}
\usepackage[applemac]{inputenc}
\usepackage[T1]{fontenc}
\pdfoutput=1
\usepackage{enumerate}
\usepackage{braket}
\usepackage{amsmath}
\usepackage{subfigure}
\usepackage{lipsum}
\usepackage{amsfonts}
\usepackage{amssymb, mathrsfs}
\usepackage{bm}
\usepackage[pdftex]{graphicx}
\usepackage{footnote}
\usepackage{hyperref}  
\hypersetup{colorlinks=true,allcolors=blue}
\usepackage{hypcap}   
\usepackage{bookmark}
\usepackage{dsfont,mathrsfs,color,url,verbatim,booktabs}
\usepackage{booktabs}
\usepackage{tabularx}

\def\beq{\begin{equation}}
	\def\eeq{\end{equation}}
\def\bsp{\begin{split}}
	\def\esp{\end{split}}
\def\bea{\begin{eqnarray}}
	\def\eea{\end{eqnarray}}
\def\ba{\begin{array}}
	\def\ea{\end{array}}

\def\l.{\left.}
\def\r.{\right.}

\def\part{\partial}

\begin{document}
\title{Conceptual and Geometric Foundations for a Teleparallel Approach to Quantum Gravity}  
\author{A. Landry}
\email{a.landry@dal.ca}
\affiliation{Department of Mathematics and Statistics, Dalhousie University, Halifax, Nova Scotia, Canada, B3H 3J5}

\begin{abstract}
{	We revisit quantum field theory in curved spacetime (QFTCS) as a semi-classical framework for quantum matter on classical geometries, emphasizing its limitations, including vacuum ambiguity and background dependence. We briefly review major approaches to { quantum gravity (QG), including Loop Quantum Gravity (LQG), string theory, and asymptotic safety, highlighting} their conceptual challenges. Motivated by these issues, we outline a teleparallel framework based on coframe and spin-connection variables, where gravity is encoded in torsion rather than curvature. This framework naturally incorporates local Lorentz symmetry and fermionic couplings while displaying a gauge-like structure. { We argue that the coframe/spin-connection pair provides an alternative and geometrically refined description of gravitational variables, which may serve as a useful starting point for future investigations of QG. The purpose of this work is not to provide a complete quantization of teleparallel gravity but to identify the geometric and conceptual ingredients that such a formulation would require.}}\\

\end{abstract}

\maketitle

\section{Introduction}

The construction of a consistent theory of quantum gravity (QG) remains one of the most profound and persistent challenges in modern theoretical physics. Despite the extraordinary empirical success of quantum field theory (QFT) in describing the Standard Model of particle physics and~the equally remarkable success of general relativity (GR) in explaining gravitational phenomena across astrophysical and cosmological scales, these two frameworks remain fundamentally incompatible at both conceptual and mathematical levels~\cite{Wald1984,Birrell1982,Kiefer2012,Weinberg1979,Crowther2025,Esposito2024,Smolin2001,Smolin2023}.

This incompatibility is not merely technical but~reflects a deep structural tension in our understanding of spacetime. In~QFT, spacetime is treated as a fixed, non-dynamical background, typically Minkowskian or weakly curved, on~which quantum fields evolve~\cite{Birrell1982,Parker2009}. In~contrast, GR describes spacetime as a dynamical entity whose geometry is determined by the distribution of matter and energy. This asymmetry becomes unavoidable at the Planck scale, where quantum fluctuations of spacetime itself are expected to be significant~\cite{Hawking1975,DeWitt1967}.

Attempts to quantize gravity perturbatively lead to a non-renormalizable theory, requiring an infinite number of counterterms~\cite{tHooft1974,Weinberg1979}. This failure suggests that GR, when treated as a QFT of the metric, is at best an effective field theory valid below the Planck scale~\cite{Donoghue1994,Burgess2004}. Moreover, conceptual problems such as the breakdown of locality, the~absence of a preferred time parameter, and~ambiguities in defining observables further complicate the formulation of a consistent theory~\cite{Kiefer2012,Weinberg1979,Crowther2025,Esposito2024,Smolin2001,Smolin2023,Isham1992}.

In recent decades, several major research programs have attempted to resolve these issues. Loop Quantum Gravity (LQG) provides a non-perturbative and background-independent quantization of geometry, predicting a fundamentally discrete structure of spacetime~\cite{Rovelli2020,Ashtekar2004}. However, its semiclassical limit and connection to low-energy physics remain under active investigation. In~addition, the~canonical formulation leads to the so-called ``problem of time'', embodied in the Wheeler--deWitt equation, which lacks an explicit temporal parameter~\cite{DeWitt1967,Kuchar1992}.

{String theory offers a distinct approach, replacing point particles with extended one-dimensional objects and naturally incorporating a massless spin-2 excitation interpreted as the graviton~\cite{Polchinski1998,Green1987}. It provides a framework capable of unifying all fundamental interactions and has led to profound insights such as holography and gauge/gravity duality~\cite{Maldacena1999}. Nevertheless, the~theory typically requires extra spatial dimensions and supersymmetry, neither of which has been experimentally confirmed. Furthermore, the~vast landscape of possible vacua limits predictive power~\cite{Susskind2003}, and~most formulations remain~background-dependent.}

The asymptotic safety program proposes that gravity may be non-perturbatively renormalizable due to the existence of an ultraviolet fixed point~\cite{Reuter2012,Percacci2017,Eichhorn2019}; while functional renormalization group techniques provide encouraging evidence, the~approach relies on truncations of theory space, and~the existence of a fully consistent continuum limit remains an open~question.

Beyond these mainstream approaches, alternative frameworks such as causal set theory, causal dynamical triangulations, and~emergent gravity models suggest that spacetime itself may not be fundamental~\cite{Sorkin2005,Loll2019,Verlinde2017,Oriti2014}. These approaches offer valuable conceptual insights into discreteness and emergence but~face significant challenges in recovering classical spacetime and connecting with observational~data.

Despite their diversity, { many approaches retain, either directly or through their classical limit, an~important relation to metric-based spacetime geometry.} This observation suggests that the central difficulty in QG may not lie solely in the quantization procedure, but~in the choice of fundamental~variables.

A deeper issue concerns the role of time. In~standard QFT, time is treated as an external parameter, whereas in GR it is dynamical and intertwined with spatial geometry. This mismatch leads to profound conceptual difficulties in defining quantum evolution and observables~\cite{Isham1992,Kuchar1992}. It is therefore natural to examine whether alternative geometric variables may provide a useful description of spacetime at the quantum~level.

{ In this work, we explore the possibility that some conceptual difficulties in QG may motivate the investigation of alternative geometric variables beyond the metric formulation, including formulations based on local frames.}

{Teleparallel gravity provides a useful alternative in this direction. In~this framework, gravitation is attributed to torsion rather than curvature, and~the { coframe and the flat spin-connection are taken as the primary variables, while the metric is reconstructed from the coframe~\cite{Aldrovandi2013,Krssak2019,Bahamonde2023,Ferraro2007,33}.} This formulation is dynamically equivalent to GR at the classical level, while naturally accommodating spinor fields and admitting a gauge-theoretic~interpretation.}

{ From this perspective, a~teleparallel formulation based on coframe/spin-connection pairs may provide an alternative geometric setting in which some conceptual issues of QG can be reconsidered. In~particular, it 	offers a framework in which torsion, local Lorentz covariance, and~frame variables are treated explicitly. At~present, there is no direct experimental evidence that spacetime torsion constitutes an independent propagating degree of freedom in nature. Whether this leads to a complete and internally consistent quantum theory remains an open~problem.

{Earlier studies of QG in torsionful geometries include renormalization analyses and QFT-type treatments in Riemann--Cartan and metric-affine backgrounds~\cite{Tseytlin1982,LeeNeeman1990,Lee1992,ShapiroTeixeira2014,Brandt2024a,Brandt2024b}. The~present work differs from these approaches by focusing specifically on the teleparallel sector, where curvature is constrained to vanish and torsion carries the gravitational~information.	}

The contribution of this paper is threefold. First, we clarify the role of coframe and spin-connection variables in teleparallel gravity as candidate variables for a quantum formulation. Second, we compare this framework with curvature-based, connection-based, and~metric-affine approaches. Third, we outline a constrained canonical setting in which the coframe is treated as the main gravitational variable while the flat spin connection encodes inertial Lorentz covariance. This paper is therefore programmatic in scope. It does not claim to solve the problem of observables, renormalizability, or~the construction of the physical
Hilbert space. Rather, it identifies the geometrical structures that must be controlled in any teleparallel QG proposal. Accordingly, the~present work should be viewed as a conceptual and geometric analysis rather than as a completed QG construction.
}

The aim of this work is therefore twofold. In~Sections~\ref{sect2} and \ref{sect3}, we provide a critical analysis of existing approaches to QG, emphasizing their structural limitations. In~Sections~\ref{sect4} and \ref{sect5}, we explore the possibility that a teleparallel, frame-based formulation { may provide a useful framework for formulating open questions about torsion-based quantum geometry.}

{
\subsection*{Notation}

Throughout this work we use the metric signature $(-,+,+,+)$ and natural units $c=\hbar=1$. Greek indices $\mu,\nu,\ldots$ denote spacetime coordinate indices, whereas Latin indices $a,b,\ldots$ denote Lorentz-frame indices. The~coframe is denoted by $h^a_{\mu}$, its inverse by $h_a^{\mu}$, and~its determinant by $h=\det(h^a_{\mu})=\sqrt{-g}$. The~inverse coframe satisfies
\begin{equation}
h^a_{\mu} h_a^{\nu}=\delta_\mu^{\nu},
\qquad
h^a_{\mu} h_b^{\mu}=\delta^a_{~b} .
\end{equation}
 The gravitational coupling is $\kappa=8\pi G$.
}

\newpage

\section{Quantum Field Theory in Curved~Spacetime}\label{sect2}
\unskip

\subsection{Klein--Gordon~Field}

The dynamics of a scalar field in curved spacetime are governed by the action
\begin{equation}
	S_\phi =-\frac{1}{2}\int d^4x\,h\left(g^{\mu\nu}\nabla_\mu\phi\nabla_\nu\phi
	+m^2\phi^2+\xi R\phi^2	\right),
\end{equation}
leading to the Klein--Gordon equation
\begin{equation}
(\Box-m^2-\xi R)\phi=0, \qquad \Box=g^{\mu\nu}\nabla_\mu\nabla_\nu .
\end{equation}
{Here, $h=\sqrt{-g}=\det(h^a{}_\mu)$ according to the convention introduced above. Physically admissible states are commonly restricted to Hadamard states in order to ensure a well-defined renormalized stress-energy tensor.}

On globally hyperbolic spacetimes, one constructs a complete set of mode solutions and defines a Hilbert space using the Klein--Gordon inner product~\cite{Fulling1989, Birrell1982}. However, the~absence of a preferred timelike Killing vector leads to ambiguities in the definition of vacuum states~\cite{Fulling1973, Wald1994}.

\subsection{Dirac~Field}

{ The coframe formalism provides the standard and geometrically transparent framework for coupling spinor fields to curved spacetime, although~alternative non-tetrad formulations have been explored}. The~Dirac action is given by
\begin{equation}
	S = \int d^4x \, h \, \bar{\psi} (i\gamma^a h_a{}^\mu D_\mu - m)\psi,
\end{equation}
where \(D_\mu\) is the spinor covariant derivative containing the spin-connection~\cite{Birrell1982,Hehl1976}. The~corresponding Dirac equation is
\begin{equation}
	(i\gamma^a h_a{}^\mu D_\mu - m)\psi=0.
\end{equation}

Quantization proceeds via a mode expansion and canonical anti-commutation relations. { The usefulness of coframes in the standard spinorial formulation highlights the central role of local Lorentz symmetry and motivates the use of 	frame-based variables in teleparallel gravity. Nonlinear geometric formulations of spinors without introducing an independent orthonormal tetrad have also been discussed in the literature, notably by 	Ogievetsky and Polubarinov and in later analyses by Pitts~\cite{OgievetskyPolubarinov,Pitts}. Thus, the~use of coframes should not be interpreted as the only possible description of spinors in curved spacetime, but~rather as the natural language for teleparallel and Lorentz-covariant formulations.
}

\subsection{Proca~Field}

The Proca field describes a massive spin-1 field with action
\begin{equation}
S_A =
\int d^4x\,h\left[-\frac{1}{4}F_{\mu\nu}F^{\mu\nu}
-\frac{1}{2}m^2 A_\mu A^\mu\right],\qquad F_{\mu\nu}=\nabla_\mu A_\nu-\nabla_\nu A_\mu ,
\end{equation}
{ where \(F_{\mu\nu}\) is the antisymmetric field strength.}

The field satisfies
\begin{equation}
	\nabla_\mu F^{\mu\nu}-m^2 A^\nu=0.
\end{equation}
{For $m\neq 0$, taking the divergence of the field equation (FEs) yields the Proca constraint $\nabla_\mu A^\mu=0$ \cite{AllenJacobson1986}.} The theory propagates three physical polarization states, reflecting the degrees of freedom of a massive vector~field.

\subsection{Spin-2 Field and Gravitational~Waves}

Linearized gravity describes perturbations \(\gamma_{\mu\nu}\) around a background metric. { Expanding the Einstein--Hilbert action to second order in the metric perturbation yields the quadratic action for the spin-2 field, together with curvature-dependent interaction terms when the background is not flat.  We decompose the metric as
\begin{equation}
	g_{\mu\nu}=\bar g_{\mu\nu}+\gamma_{\mu\nu},
\end{equation}
where $\bar g_{\mu\nu}$ is the background metric. The~trace-reversed
perturbation is
\begin{equation}
\bar{\gamma}_{\mu\nu} = \gamma_{\mu\nu}-\frac{1}{2}\bar g_{\mu\nu}\gamma,
\qquad
\gamma=\bar g^{\mu\nu}\gamma_{\mu\nu}.
\end{equation}
In curved spacetime, the~FEs take the form
\begin{equation}
	\Box \bar{\gamma}_{\mu\nu} + 2R_{\mu\alpha\nu\beta}\bar{\gamma}^{\alpha\beta} = 0,
\end{equation}
 where \(\Box=\bar g^{\mu\nu}\bar\nabla_\mu\bar\nabla_\nu\) is the d'Alembertian associated with \(\bar g_{\mu\nu}\).

Although a graviton interpretation can be introduced perturbatively around suitable backgrounds}, perturbative quantization leads to non-renormalizable divergences~\cite{tHooft1974, GoroffSagnotti1986}. This indicates that the metric-based spin-2 field theory is only an effective description valid at low energies~\cite{Donoghue1994}.

\subsection{Critical~Assessment}

Quantum field theory in curved spacetime (QFTCS) provides a consistent and mathematically well-defined framework for describing quantum matter propagating on classical gravitational backgrounds. It successfully accounts for key physical effects such as particle creation in expanding universes~\cite{Parker1969} and black hole evaporation~\cite{Hawking1975} and~offers a rigorous treatment of renormalized observables through the use of Hadamard states and local covariance~\cite{Wald1994, Birrell1982}. However, the~theory is intrinsically semi-classical: matter fields are quantized while spacetime geometry remains classical, leading to an inherent inconsistency when backreaction effects become significant~\cite{Birrell1982}. This limitation is particularly evident in regimes where quantum fluctuations of the gravitational field cannot be~neglected.

{
The problem of observables is not unique to metric formulations and may arise in any generally covariant theory. The~teleparallel framework does not by itself solve this problem; rather, it provides a different set of variables in which the issue may be reconsidered.
}

A further conceptual difficulty arises from the absence of a preferred vacuum state in generic curved spacetimes. In~the absence of a global timelike Killing vector, the~notion of particles becomes observer-dependent, as~illustrated by inequivalent quantizations associated with different mode decompositions~\cite{Fulling1973, Wald1994}. Moreover, the~formalism relies on a fixed background geometry, which stands in tension with the dynamical nature of spacetime in GR, while QFTCS can be interpreted as an effective field theory valid at energies below the Planck scale~\cite{Donoghue1994}; it does not by itself provide a complete theory of quantum gravity. { These limitations motivate the search for formulations in which the geometrical
degrees of freedom may be organized differently at the quantum level.}

\section{Geometric Variables in Quantum Gravity~Approaches}\label{sect3}

A wide range of approaches have been developed to reconcile quantum mechanics (QM) and GR; while each provides important insights, none has yet achieved a complete and experimentally verified~theory.

\subsection{Loop Quantum~Gravity}

LQG is a non-perturbative and background-independent approach to QG based on the canonical quantization of GR in terms of connection variables~\cite{Ashtekar1986, Rovelli2004, Thiemann2007}.

The fundamental phase space variables are the $\mathrm{SU}(2)$ connection $A^i_A$ and the densitized triad $E^A_i$, satisfying
\begin{equation}
	\{A^i_A, E^B_j\} = 8\pi G \gamma \delta^B_A \delta^i_j \delta(x,y),
\end{equation}
where $\gamma$ is the Barbero--Immirzi parameter. { Here, $A,B,\ldots$ denote spatial indices and $i,j,\ldots$ internal $SU(2)$ indices.}

Quantum states are represented by spin networks, which form an orthonormal basis of the kinematical Hilbert space~\cite{Rovelli2004, Ashtekar2004}. These states diagonalize geometric operators such as area and volume, which possess discrete~spectra.

The dynamics are governed by the Hamiltonian constraint, leading to the Wheeler--deWitt equation defined by~\cite{Thiemann2007}:
\begin{equation}
	\hat{H}\Psi = 0,
\end{equation}
which encodes the dynamics of quantum geometry. { However, explicit solutions remain difficult to construct, and~the implementation
of the Hamiltonian constraint remains technically nontrivial and not uniquely defined.}

A covariant formulation is provided by spin foam models, which define transition amplitudes between spin network states~\cite{Perez2013}. In~symmetry-reduced settings, loop quantum cosmology predicts a resolution of the Big Bang singularity via a quantum bounce~\cite{Ashtekar2006}.

Although both LQG and teleparallel gravity use frame-related variables, their geometric interpretations differ. LQG is based on the Ashtekar--Barbero $SU(2)$ connection and densitized triad, with~curvature holonomies playing a central role. Thus, the~distinction is not merely between metric and frame variables, but~between curvature-based and torsion-based descriptions of gravitational field strength. Teleparallel gravity instead imposes vanishing curvature for the Weitzenb\"ock connection and attributes gravity to torsion. This does not mean that the Levi--Civita curvature associated with the metric vanishes; rather, the~curvature of the teleparallel spin connection is set to zero, while the Levi--Civita curvature is encoded equivalently through torsion and a boundary term.

Despite its conceptual strengths, LQG faces challenges, including ambiguities in the dynamics, the~problem of time, and~the recovery of classical spacetime~geometry.

\subsection{String~Theory}

{ String theory provides a distinct route to quantum gravity by replacing point particles with one-dimensional extended objects whose 	perturbative dynamics are described by the Polyakov action~\cite{Polchinski1998,Green1987}. Upon~quantization, the~closed-string spectrum contains a massless spin-2 excitation, naturally interpreted as the~graviton.
	
	For the purposes of the present work, the~main lesson of string theory is that 	the microscopic description of gravity may require structures different from those used in classical GR. However, in~most perturbative formulations, the 	theory is developed around a chosen background geometry, and~the metric remains central in the low-energy effective description~\cite{Polchinski1998,DouglasKachru2007}. Dualities and the AdS/CFT correspondence provide powerful non-perturbative insights~\cite{Maldacena1999}, but~the relation to a manifestly background-independent formulation of spacetime geometry remains subtle. Moreover, the~non-perturbative definition of the full theory remains incomplete, in general,~backgrounds.
	
	String theory therefore serves here as a contrast with the teleparallel approach: string theory modifies the microscopic objects, whereas teleparallel gravity modifies the geometrical variables used to describe the gravitational
	field.
}

\subsection{Asymptotic~Safety}

The asymptotic safety program proposes that QG may be defined as a non-perturbatively renormalizable QFT through the existence of a non-trivial ultraviolet (UV) fixed point of the renormalization group flow~\cite{tHooft1974, Donoghue1994,Reuter2012,Eichhorn2019,Percacci2017}. In~this framework, the~effective gravitational action is described by a scale-dependent functional whose couplings approach finite values at high energies, ensuring predictivity despite the perturbative non-renormalizability of Einstein gravity. Functional renormalization group (FRG) techniques have provided evidence for such fixed points in truncated theory spaces, typically involving higher-curvature operators such as $R^2$ and $R_{\mu\nu}R^{\mu\nu}$ terms~\cite{Donoghue1994}. 

Despite these encouraging results, the~asymptotic safety scenario remains subject to important limitations. In~particular, the~existence and properties of the UV fixed point depend on truncation schemes, raising questions about robustness and convergence. Moreover, many formulations are expressed in terms of the metric field, sharing some conceptual issues of perturbative approaches, including background dependence and difficulties in defining physical observables. { The extent to which asymptotic safety can be formulated in a fully background-independent manner remains an active area of research. For~the present comparison, its relevance lies in the fact that many standard formulations are metric-based, although~tetrad and connection versions have also been studied. In~this sense, coframe- and torsion-based formulations provide a useful comparison point for assessing the role of geometric~variables.

\subsection{Einstein--Cartan and Metric-Affine~Approaches}

The teleparallel framework should also be situated within the broader class of non-Riemannian geometrical formulations of gravity. In~Einstein--Cartan theory, the~affine connection may have torsion in addition to curvature, and~torsion is algebraically related to the spin density of matter~\cite{Hehl1976}. In~this setting, torsion does not replace curvature as the carrier of gravity; rather, curvature and torsion coexist as geometrical structures associated with the
connection.

Metric-affine gravity provides a still more general framework in which the
metric and affine connection are treated as independent variables. The~corresponding geometry may possess curvature, torsion, and~nonmetricity as
independent field strengths~{\cite{Hehl1976,ObukhovPereira2003}.} 
 Teleparallel gravity corresponds to the sector in which the curvature of the teleparallel connection and the nonmetricity vanish, while torsion remains nonzero and encodes the gravitational interaction. The~Levi--Civita curvature associated with the reconstructed metric need not~vanish.

This distinction is important because earlier quantum studies involving
torsion often belong to the Einstein--Cartan or metric-affine setting rather than to the teleparallel sector considered here
~\cite{Tseytlin1982,LeeNeeman1990,Lee1992,ShapiroTeixeira2014,Brandt2024a,Brandt2024b}. This clarifies why the present proposal is not a generic quantization of torsion, but~a teleparallel one in which the flat spin connection plays a specific inertial~role.

\subsection{Alternative and Emergent~Approaches}

Several approaches to quantum gravity question whether smooth spacetime geometry is the most primitive level of description. Causal set theory, causal dynamical triangulations, group field theory, induced gravity, entropic gravity, and~holographic scenarios suggest, in~different ways, that the continuum spacetime of GR may arise from more primitive structures or collective degrees of freedom~\cite{Sorkin2005,Loll2019,Oriti2014,Verlinde2017,Maldacena1999}.

The relevance of these approaches for the present work is mainly conceptual. They show that the metric formulation of GR need not be the only starting point for thinking about quantum spacetime. However, the~teleparallel proposal considered here is different from emergent-spacetime scenarios: it retains a differentiable spacetime manifold but reorganizes the gravitational variables in terms of a coframe and a flat spin-connection, with~torsion playing the role of gravitational field~strength.


\subsection{{Comparative~Summary and Critical~Analysis} 
}
\vspace{-6pt}

\begin{table}[ht]
\caption{{Comparison of geometrical} 
 variables and field strengths in selected approaches to quantum~gravity.}
	\label{tab:variables}
	\centering
\fontsize{8.5}{8.5}\selectfont
	
\begin{tabularx}{\textwidth}{llll}
		\toprule
		\textbf{Approach} & \textbf{Basic Variables} & \textbf{Field Strength} & \textbf{Main Open Issue} \\
		\midrule
		Metric GR/QFT & $g_{\mu\nu}$ & Levi--Civita curvature &
		Perturb.  nonrenormalizability \\
		\midrule
		LQG & $A^i_A$, $E^A_i$ & Curvature holonomies &
		Semiclassical limit \\
		\midrule
		String theory & Worldsheet fields, background metric &
		Spin-2 excitation & Vacuum/background selection \\
		\midrule
		Einstein--Cartan & Coframe, Lorentz connection &
		Curvature and torsion & Spin--torsion sector \\
		\midrule
		Metric-affine gravity & Metric, affine connection  &
		Curvature, torsion,  nonmetricity & Many degrees of freedom \\
		\midrule
		Teleparallel gravity  & $h^a{}_\mu$, $\omega^a{}_{b\mu}$ &
		Torsion & Quantum formulation open \\
		\bottomrule
	\end{tabularx}

\end{table}
\normalsize



The comparison as presented in table \ref{tab:variables} suggests that different quantum-gravity programs are
distinguished not only by their quantization methods but~also by their choice of geometrical variables. Metric, connection, coframe, torsion, nonmetricity, and~extended objects emphasize different aspects of the gravitational~field.

This observation should be interpreted cautiously. The~use of coframes and
torsion does not by itself solve the problems of observables, dynamics,
renormalizability, or~the recovery of the classical limit. In~particular, the~conceptual issues associated with diffeomorphism invariance and constrained Hamiltonian systems may reappear in any generally covariant theory, including a teleparallel~one.

The motivation for the teleparallel approach is therefore specific and deliberately limited: it provides a geometrical setting in which the metric is reconstructed from the coframe, the~spin-connection is flat and inertial, and~torsion carries the gravitational information. Whether this variable choice leads to a consistent quantum theory remains an open question. The~next section therefore turns to the teleparallel case in more detail, emphasizing the precise geometrical meaning of the coframe, flat spin-connection, and~torsion~scalar.

}

\newpage

\section{Teleparallel Gravity within the Landscape of Non-Riemannian~Geometries}\label{sect4}

Teleparallel gravity provides an alternative geometric formulation of gravitation in which torsion, rather than curvature, encodes the gravitational interaction~\cite{Aldrovandi2013,21}. The~fundamental variables are the coframe $h^a_{\ \mu}$ and spin-connection $\omega^a_{~b\mu}$, which define a local orthonormal basis on spacetime and reconstruct the metric via:
\begin{equation}
	g_{\mu\nu} = \eta_{ab} \, h^a_{\ \mu} h^b_{\ \nu} \quad\text{and}\quad \omega^a{}_{b\mu}	=
	\Lambda^a{}_c \partial_\mu (\Lambda^{-1})^c{}_b ,
\end{equation}
where \(\Lambda^a{}_b(x)\) is a local Lorentz transformation specifying the inertial~spin-connection.

\subsection{Geometric~Foundations}

The theory is based on the Weitzenb\"ock connection:
\begin{equation}
\Gamma^\lambda{}_{\nu\mu} = h_a{}^\lambda\left(\partial_\mu h^a{}_\nu+\omega^a{}_{b\mu}h^b{}_\nu\right),
\end{equation}
which is curvature-free but torsionful:
\begin{align}
	T^a{}_{\mu\nu}	=&	\partial_\mu h^a{}_\nu-\partial_\nu h^a{}_\mu	+\omega^a{}_{b\mu}h^b{}_\nu	-\omega^a{}_{b\nu}h^b{}_\mu ,
		\\
	T^{\lambda}_{\mu\nu} =& \Gamma^{\lambda}_{\ \nu\mu} - \Gamma^{\lambda}_{\ \mu\nu},
	\\
	S_\lambda{}^{\mu\nu}=& \frac12 \left(
	K^{\mu\nu}{}_\lambda +\delta^\mu_\lambda T^{\alpha\nu}{}_\alpha
	-\delta^\nu_\lambda T^{\alpha\mu}{}_\alpha
	\right).
\end{align}
{Here, \(K^\lambda{}_{\mu\nu}\) is the contorsion tensor, relating the Weitzenb\"ock and Levi--Civita connections, and~is defined by}
\begin{equation}
K^\lambda{}_{\mu\nu} =\frac{1}{2}\left(T_\mu{}^\lambda{}_\nu+T_\nu{}^\lambda{}_\mu-T^\lambda{}_{\mu\nu}\right).
\end{equation}

The torsion scalar is constructed as:
\begin{equation}
	T = \frac{1}{4} T^{\lambda}_{\ \mu\nu} T_{\lambda}^{\ \mu\nu}
	+ \frac{1}{2} T^{\lambda}_{\ \mu\nu} T^{\nu\mu}_{\ \ \lambda}
	- T^{\lambda}_{\ \mu\lambda} T^{\nu\mu}_{\ \ \nu}.
\end{equation}

The TEGR action reads:
\begin{equation}
	S_{\mathrm{TEGR}} = \frac{1}{2\kappa} \int d^4x \, h \, T,
\end{equation}
and is dynamically equivalent to GR provided appropriate boundary conditions are imposed~\cite{Aldrovandi2013,21}. { The equivalence with GR follows from the identity \(\overset{\circ}{R}=-T+B\), where \(B\) is a boundary~term.

}

\subsection{Extension to \texorpdfstring{$F(T)$}{F(T)} Gravity}

A natural modification consists of promoting the torsion scalar to a function:
\begin{equation}
	S = \frac{1}{2\kappa} \int d^4x \, h \, F(T),
\end{equation}
leading to modified gravitational dynamics~\cite{22,Bahamonde2023,25,ColeyLandry2024}. Variation of the action gives the symmetric and antisymmetric parts of the FEs:
\begin{eqnarray}
	\kappa\,\Theta_{\left(ab\right)} &=& F_T\left(T\right) \overset{\ \circ}{G}_{ab}+F_{TT}\left(T\right)\,S_{\left(ab\right)}^{~~\mu}\,\partial_{\mu} T+\frac{g_{ab}}{2}\,\left[F\left(T\right)-T\,F_T\left(T\right)\right],  \label{1001a}
	\\
	0 &=& F_{TT}\left(T\right)\,S_{\left[ab\right]}^{~~\mu}\,\partial_{\mu} T, \label{1001b}
\end{eqnarray}
{Here, \(F_T=\frac{dF}{dT}\), \(F_{TT}=\frac{d^2F}{dT^2}\), \(\Theta_{ab}\) is the matter energy-momentum tensor, \(\overset{\circ}{G}_{ab}\) is the Einstein tensor
	computed with the Levi--Civita connection, and~\(S_a{}^{\mu\nu}\) is the teleparallel superpotential.} Adding the GR conservation laws $\overset{\ \circ}{\nabla}_{\nu}\left(\Theta^{\mu\nu}\right)=0$, { the antisymmetric equation imposes compatibility conditions on admissible coframe/spin-connection pairs in \(F(T)\) models.}  In the covariant formulation, the~inclusion of a flat spin connection restores local Lorentz covariance and separates inertial effects from genuine
	gravitational torsion. In~contrast to \(F(R)\) theories, the~FEs remain second order, which avoids Ostrogradsky instabilities and represents a key structural~advantage.

The role of the spin connection is central in the covariant formulation of \(F(T)\) gravity. A~major conceptual issue in early $F(T)$ formulations was the apparent violation of local Lorentz invariance. This problem is resolved in the covariant approach developed by Kr\v{s}\v{s}\'ak and Pereira~\cite{24}, where the coframe $h^a_{\ \mu}$ and a flat spin-connection $\omega^a_{\ b\mu}$ are treated as independent variables. This formulation ensures proper separation between inertial and gravitational effects and restores local Lorentz covariance at the level of the action and FEs. { However, modified teleparallel theories may introduce additional propagating degrees of freedom whose physical interpretation remains under investigation.}

\subsection{Recent Developments and Exact~Solutions}

A significant body of recent work has focused on constructing exact solutions and exploring the phenomenology of $F(T)$ gravity.

{ Recent studies have constructed several classes of exact solutions in teleparallel $F(T)$ gravity, including static spherically symmetric perfect-fluid configurations and anisotropic cosmological models~\cite{Landry2024,Landry2025}.} These studies demonstrate that torsion-based dynamics can reproduce a wide range of physically relevant spacetimes while introducing novel structural features absent in curvature-based theories. { These constructions illustrate how torsion-based variables can encode nontrivial gravitational configurations in modified teleparallel models.}

Complementary studies, including those by Sahoo and collaborators, have investigated cosmological constraints and late-time acceleration scenarios within $F(T)$ models~\cite{26}, while dynamical systems analyses by Coley and van den Hoogen emphasize the role of geometric variables in determining the global structure of cosmological phase space~\cite{27}. { Several recent reviews and phenomenological studies have also emphasized the role of modified teleparallel models in cosmology, compact objects, perturbation theory, and~observational constraints~\cite{Bahamonde2023,25}.}

\subsection{New GR and Extended Teleparallel~Theories}

Beyond TEGR and its \(F(T)\) extensions, a~broader class of torsion-based theories has been developed under the framework of New General Relativity (NGR). In~this approach, the~gravitational Lagrangian is constructed from the three irreducible quadratic invariants of the torsion tensor:
\begin{equation}
	\mathcal{L}_{\mathrm{NGR}} = a_1 T^{\lambda}_{\ \mu\nu} T_{\lambda}^{\ \mu\nu}
	+ a_2 T^{\lambda}_{\ \mu\nu} T^{\nu\mu}_{\ \ \lambda}
	+ a_3 T^{\lambda}_{\ \mu\lambda} T^{\nu\mu}_{\ \ \nu},
\end{equation}
where $a_1$, $a_2$, and~$a_3$ are free parameters~\cite{34,35}.

TEGR corresponds to a specific choice of these coefficients, while NGR allows for a wider parameter space of torsion-based dynamics. This generalization provides a useful testing ground for exploring deviations from GR within a purely torsional framework. { NGR may be viewed as a torsion-quadratic sector within the broader landscape of gravitational gauge theories. It remains distinct from Einstein--Cartan and metric-affine models because curvature and nonmetricity are not treated as independent gravitational \mbox{field strengths.}}

However, NGR models typically introduce additional propagating degrees of freedom, and~their physical viability depends sensitively on parameter choices. Stability, ghost freedom, and~consistency with observations impose strong constraints, limiting the range of acceptable~models.

\subsection{Boundary-Term Extensions: \texorpdfstring{$F(T,B)$}{F(T,B)} Gravity}

An important extension of teleparallel gravity involves the boundary term $B$, defined through the relation:{
\begin{align}
	R=-T+B,\qquad
	B=\frac{2}{h}\partial_\mu(hT^\mu),
	\qquad
	T^\mu=T^\nu{}_\nu{}^\mu ,
\end{align}
which connects the Ricci scalar $R$ to the torsion scalar $T$.}

This leads to $F(T,B)$ gravity, with~action:
\begin{equation}
	S = \frac{1}{2\kappa} \int d^4x \, h \, F(T,B),
\end{equation}
which interpolates between $F(T)$ and $F(R)$ theories~\cite{33,Bahamonde2023}.

Unlike pure $F(T)$ models, $F(T,B)$ theories generally yield fourth-order FEs due to the presence of higher derivatives through $B$. As~a result, they reintroduce some of the complexities associated with curvature-based modified gravity, including potential~instabilities.

Nevertheless, $F(T,B)$ gravity provides a unifying framework that clarifies the relationship between torsion and curvature formulations and highlights the role of boundary terms in gravitational~dynamics.

\subsection{Nonmetricity and Hybrid Extensions: \texorpdfstring{$F(Q)$}{F(Q)} and \texorpdfstring{$F(T,Q)$}{F(T,Q)} Theories}

A further generalization arises in the context of symmetric teleparallel gravity, where gravity is attributed to nonmetricity rather than torsion or curvature~\cite{38}. In~this framework, the~fundamental scalar is the nonmetricity scalar $Q$, leading to $F(Q)$ theories. {The nonmetricity tensor is
\begin{equation}
	Q_{\alpha\mu\nu}=\nabla_\alpha g_{\mu\nu},
\end{equation}
and \(Q\) denotes the corresponding quadratic nonmetricity scalar, up~to sign conventions. Thus, \(F(Q)\) models belong to the symmetric teleparallel sector, whereas \(F(T)\) models belong to the torsional teleparallel sector.}

{More recently, hybrid extensions combining torsion and nonmetricity, such as $F(T,Q)$ models, have been proposed as part of a broader unification of geometric formulations~\cite{39}. These theories suggest that curvature, torsion, and~nonmetricity may represent different aspects of a deeper underlying geometric structure. However, they also significantly enlarge the space of possible models, raising concerns about predictability and physical~interpretability.}

\subsection{Conceptual and Critical Implications for Quantum~Gravity}

The teleparallel and coframe-based formulation of gravity offers a conceptually distinct perspective in which the fundamental variables are the coframe fields and spin--connection, rather than the metric tensor. In~this framework, gravitation is attributed to torsion rather than curvature, and~the dynamical structure resembles that of a gauge theory for translations~\cite{Hehl1976}. This shift in geometric interpretation is relevant for QG. In~particular, the~coframe formalism naturally accommodates fermionic fields and local Lorentz symmetry, addressing limitations already encountered in QFTCS~\cite{Birrell1982}. Moreover, the~analogy between torsion and field strength suggests that gravitational interactions may potentially be reformulated in a way partly analogous to quantization procedures used in gauge~theories.

From a conceptual standpoint, the~teleparallel approach motivates reconsidering the role of the metric as the fundamental descriptor of spacetime geometry. As~highlighted in various QG programs, including LQG and effective field theory approaches~\cite{Rovelli2004, Donoghue1994}, the~identification of the true dynamical degrees of freedom remains an open problem. In~this context, the~coframe and spin-connection variables provide a different geometric structure, potentially capturing both translational and rotational aspects of spacetime symmetries. This viewpoint is further supported by the natural role of coframes in the standard spinorial formulation and by the gauge-like structure underlying teleparallel gravity, which may provide a useful alternative setting for quantization relative to purely metric-based~formulations.

Despite these promising features, several challenges remain. The~role of torsion at the quantum level, the~construction of a consistent quantum theory based on coframe variables, and~the recovery of classical GR in an appropriate limit all require further investigation. Additionally, the~relation between teleparallel gravity and other approaches, such as string theory or asymptotic safety, is not yet fully understood~\cite{Polchinski1998, Donoghue1994}. { Nevertheless, the~teleparallel framework provides a complementary geometric setting in which some conceptual issues can be reformulated. It thus represents a possible direction for future investigations of torsion-based quantum geometry.}

\section{ Toward a Constrained Quantum Teleparallel~Framework}\label{sect5}

{ The construction below is programmatic, formal, and~heuristic in nature. It is not intended as a complete Dirac quantization of teleparallel gravity, but~as an identification of the canonical variables, constraints, and~consistency conditions that such a quantization would have to address.} In the covariant formulation of teleparallel gravity, the~gravitational degrees of freedom are encoded in the coframe field $h^a_{\ \mu}$, while the flat spin--connection $\omega^a_{\ b\mu}$ accounts for inertial effects and ensures local Lorentz invariance~\cite{Aldrovandi2013,21,24}. { Any candidate quantum formulation must specify how this constrained pair is treated.}

\subsection{Canonical Variables and Extended Phase~Space}

We perform a $3+1$ decomposition of spacetime, defining spatial coframes $h^a_{\ i}$ and their conjugate momenta $\pi_a^{\ i}$. The~spin--connection is included as a constrained variable with conjugate momentum $\Pi_a^{\ bi}$:
\begin{equation}
	(h^a_{\ i}, \pi_a^{\ i}), \quad (\omega^a_{\ bi}, \Pi_a^{\ bi} \approx 0),
\end{equation}
{where \(i,j,\ldots\) denote spatial indices on the chosen hypersurface. The~pair \((\omega^a{}_{bi},\Pi^{bi}{}_a)\) is introduced in an extended phase-space sense. Since the spin connection is flat and inertial, its conjugate momentum defines a primary constraint rather than an independent propagating degree of freedom.} The vanishing of $\Pi_a^{\ bi}$ reflects the fact that the spin--connection is non-dynamical, constrained by the flatness condition:
\begin{equation}
	R^a_{\ bij}(\omega) = \partial_i \omega^a_{\ bj} - \partial_j \omega^a_{\ bi} + \omega^a_{\ ci} \omega^c_{\ bj} - \omega^a_{\ cj} \omega^c_{\ bi} = 0.
\end{equation}

Classically, these constraints ensure that $\omega^a_{\ b\mu}$ encodes only inertial effects, preserving the covariant structure of the theory~\cite{24,32}. { This canonical decomposition introduces a foliation of spacetime and is therefore not manifestly covariant, although~it is useful for identifying the constraint structure.}

\subsection{Canonical~Quantization}

Promoting the classical fields to operators acting on a Hilbert space $\mathcal{H}$, we impose canonical commutation relations:
\begin{align}
	[\hat{h}^a_{\ i}(x), \hat{\pi}_b^{\ j}(y)] &= i\hbar \, \delta^a_b \delta_i^j \delta^{(3)}(x-y), \\
	[\hat{\omega}^a_{\ bi}(x), \hat{\Pi}_c^{\ dj}(y)] &= i\hbar \, \delta^a_c \delta^d_b \delta_i^j \delta^{(3)}(x-y),
\end{align}
with all other commutators vanishing. The~physical states $|\Psi\rangle$ satisfy the operator constraints:
\begin{equation}
	\hat{\Pi}_a^{\ bi} |\Psi\rangle = 0, \quad \hat{R}^a_{\ bij}(\omega) |\Psi\rangle = 0,
\end{equation}
which enforce the flatness of the spin connection and restrict the quantum states to the physically admissible sector. {  These commutation relations are formal and should be understood prior to 	solving the constraints. A~full quantum theory would require the construction of a physical Hilbert space after imposing all first-class constraints. Issues related to operator ordering, regularization, and~the definition of an appropriate inner product are left open.}

\subsection{Torsion Operator and Gauge~Structure}

The torsion operator is defined as
\begin{equation}
	\hat{T}^a_{\ \mu\nu} = \partial_\mu \hat{h}^a_{\ \nu} - \partial_\nu \hat{h}^a_{\ \mu}
	+ \hat{\omega}^a_{\ b\mu} \hat{h}^b_{\ \nu} - \hat{\omega}^a_{\ b\nu} \hat{h}^b_{\ \mu}.
\end{equation}

It acts as a quantum field strength associated with translational gauge symmetry, with~the coframe playing the role of a gauge potential. { This analogy with Yang--Mills theory should be interpreted geometrically rather than as a literal dynamical equivalence with an internal gauge theory. In~teleparallel gravity, the~gauge-like structure is tied to spacetime translations and local Lorentz covariance.} The spin--connection ensures that torsion transforms covariantly under local Lorentz transformations:
\begin{align}
	\delta \hat{h}^a_{\ \mu} &= \epsilon^a_{\ b}(x) \, \hat{h}^b_{\ \mu}, \\
	\delta \hat{\omega}^a_{\ b\mu} &= - D_\mu \epsilon^a_{\ b}(x),
\end{align}
where $D_\mu$ is the covariant derivative with respect to $\omega$ { and \(\epsilon^a{}_b=-\epsilon_b{}^a\) is the infinitesimal Lorentz parameter. The~generator of local Lorentz transformations is denoted by \(\hat J_{ab}\).} Physical states satisfy
\begin{equation}
	\hat{\mathcal{J}}_{ab} |\Psi\rangle = 0,
\end{equation}
implementing Lorentz invariance at the quantum level~\cite{24,32}.

\subsection{Constraint Algebra and~Dynamics}

The quantum Hamiltonian $\hat{\mathcal{H}}$, diffeomorphism generators $\hat{\mathcal{D}}_i$, and~Lorentz generators $\hat{\mathcal{J}}_{ab}$ act on $\mathcal{H}$ as:
\begin{equation}
	\hat{\mathcal{H}} |\Psi\rangle = 0, \quad
	\hat{\mathcal{D}}_i |\Psi\rangle = 0, \quad
	\hat{\mathcal{J}}_{ab} |\Psi\rangle = 0.
\end{equation}

{ Closure of the full constraint algebra is a necessary consistency condition. It is not proven here and remains an important task for future work. The~present analysis should therefore be understood as identifying the 	relevant constraint structure rather than completing the Dirac quantization program~\cite{Aldrovandi2013,21,32}. Only after such closure is established can one claim consistency of the quantum dynamics. Possible anomalies in the quantum constraint algebra remain to be~investigated.

\subsection{Comparison with Ashtekar--Barbero~Variables}

Although both approaches use frame-related variables, the~teleparallel
coframe/spin-connection pair differs from the Ashtekar--Barbero variables used in LQG. In~LQG, the~connection is an \(SU(2)\) connection whose holonomies encode curvature. In~the teleparallel setting, the~spin-connection is flat and inertial, while the torsion of the coframe encodes the gravitational field. Thus, the~distinction is not merely one of notation but~one of geometrical field strength: curvature in LQG versus torsion in teleparallel gravity. This comparison clarifies that the teleparallel proposal is not merely a rewriting of LQG variables, since the two frameworks organize gravitational geometry through distinct field~strengths.

}

\subsection{Physical Interpretation and~Outlook}

{ The discussion above suggests that coframe/spin-connection variables provide a coherent geometrical language for formulating a teleparallel quantum-gravity program. This does not establish a complete quantum theory but~identifies the variables and constraints that would have to be controlled.} In contrast to metric-based approaches, the~coframe formalism naturally incorporates local Lorentz symmetry and allows for a direct coupling to fermionic matter fields, a~feature already essential in QFTCS~\cite{Birrell1982, Hehl1976}. Moreover, the~teleparallel interpretation of gravity, in~which torsion replaces curvature as the mediator of gravitational interactions, offers a gauge-like structure that is structurally reminiscent of gauge-theoretic descriptions of other interactions. This perspective supports the idea that { the gravitational degrees of freedom may be organized in terms of local frame variables rather than only through the metric tensor. }

From a broader viewpoint, these findings resonate with key insights from existing approaches to QG. LQG emphasizes connection variables and discrete geometric structures~\cite{Rovelli2004, Thiemann2007}, while effective field theory approaches highlight the limitations of perturbative quantization based on the metric~\cite{Donoghue1994}. Similarly, string theory incorporates gravity through extended objects but typically relies on a fixed background geometry~\cite{Polchinski1998}. In~comparison, the~coframe/teleparallel framework offers a different geometrical organization of the gravitational variables. However, the~precise relation between torsion-based dynamics and these established frameworks remains an open question, and~further work is required to clarify whether these approaches are complementary or fundamentally~distinct.

Looking forward, several important directions emerge. A~key challenge is the construction of a consistent quantum theory based on coframe and spin-connection variables, including the identification of appropriate observables and the treatment of quantum fluctuations of torsion. Additionally, the~recovery of classical GR and standard cosmology in the appropriate limit must be established rigorously. From~a phenomenological perspective, it is essential to explore potential observational signatures, such as deviations from GR or novel coupling effects involving spin and torsion. {Ultimately, the~viability of this program will depend on whether it can produce a well-defined physical Hilbert space, a~closed anomaly free constraint algebra, a~controlled semiclassical limit, and~empirically \mbox{relevant predictions.}}

\section{Discussion and~Outlook}\label{sect6}

The analysis presented in this work highlights the conceptual and structural limitations of QFTCS and several principal approaches to QG. As~discussed in Section~\ref{sect2}, QFTCS provides a robust semi-classical framework for describing quantum matter on classical geometries, successfully accounting for phenomena such as particle creation and Hawking radiation~\cite{Birrell1982, Hawking1975}. However, its intrinsic background dependence and the ambiguity in the definition of vacuum states~\cite{Fulling1973, Wald1994} prevent it from offering a fully consistent description of quantum spacetime. {These shortcomings motivate the exploration of alternative formulations in 	which the geometrical degrees of freedom are treated dynamically and may be subject to quantization.}

The comparison of different QG programs in Section~\ref{sect3} further underscores the absence of a universally satisfactory framework. LQG provides a background-independent quantization of geometry in terms of connection variables, leading to discrete spectra for geometric operators~\cite{Rovelli2004, Thiemann2007}, yet faces challenges in defining dynamics and recovering the classical limit. String theory achieves a perturbatively consistent unification of interactions and naturally incorporates gravity via a spin-2 excitation~\cite{Polchinski1998, Maldacena1999}, but~relies on higher-dimensional backgrounds and exhibits limited predictive power due to the landscape of vacua. Asymptotic safety suggests a non-perturbative ultraviolet completion of gravity~\cite{Donoghue1994}, while alternative and emergent approaches propose that spacetime itself may not be fundamental~\cite{Wald1994}. { Despite their diversity, these frameworks leave open the question of which geometrical variables provide the most useful description of gravitational degrees of freedom in the quantum~regime.

 The teleparallel and coframe-based formulation discussed in Section~\ref{sect4} offers a distinct geometric perspective in which gravity is described by torsion rather than by the curvature of the Levi--Civita connection and~in which the metric is reconstructed from the coframe~\cite{Hehl1976}. This framework naturally accommodates 	local Lorentz covariance and the standard coupling of spinorial matter. As~emphasized in Section~\ref{sect5}, the~coframe/spin-connection pair provides a structured set of geometric variables for formulating a constrained quantum-gravity program, while leaving open the construction of a complete quantum~theory.

Looking forward, several open problems remain. A~key challenge is the
construction of a consistent quantum theory based on coframe and spin-connection variables, including the identification of physical observables, the~treatment of quantum torsion, and~the control of the constraint algebra. Establishing the precise relation between teleparallel gravity and other QG approaches, such as LQG, string theory, and~asymptotic safety, remains an important direction for future research. From~a phenomenological standpoint, it is also essential to investigate whether torsion-based dynamics could lead to observable deviations from GR or novel spin--torsion effects. Overall, the~teleparallel/coframe framework should be understood as a well-motivated
geometrical setting for future investigations, whose viability will depend on mathematical consistency, a~controlled semiclassical limit, and~empirical relevance.
}

\vspace{6pt} 





\section*{Acknowledgments}

{Thanks to Dr. Alan A. Coley for his advised opinion and his additional insights. Thanks also to the Atlantic General Relativity community for the presentation of the current paper materials.}




\begin{thebibliography}{999}
	
	
	
\bibitem{Birrell1982} Birrell, N.D.; Davies, P.C.W. \textit{Quantum Fields in Curved Space}; Cambridge University Press: Cambridge, UK, 1982.
	
	
\bibitem{Kiefer2012} Kiefer, C. \textit{Quantum Gravity}; Oxford University Press: Oxford, UK, 2012.
	
	
\bibitem{Weinberg1979} Weinberg, S. Ultraviolet Divergences in Quantum Theories of Gravitation. In \textit{General Relativity: An Einstein Centenary Survey}; Hawking, S.W., Israel, W., Eds.; Cambridge University Press: Cambridge, UK, 1979; pp. 790--831.
	
	
\bibitem{Crowther2025} Crowther, K. \textit{Why Do We Want a Theory of Quantum Gravity?}; Cambridge University Press:  {Cambridge, UK,} 
 2025.
	
	
\bibitem{Esposito2024} Esposito, G. {An introduction to quantum gravity}. \textit{arXiv} \textbf{2011}, arXiv:1108.3269.
	
	
\bibitem{Smolin2001} Smolin, L. \textit{Three Roads to Quantum Gravity}; Basic Books: New York, NY, USA, 2001.
	
	
\bibitem{Smolin2023} Smolin, L. The Status of Quantum Gravity. \textit{Int. J. Mod. Phys. D} \textbf{2023}, \textit{32}, 2330002.
	
	
\bibitem{Wald1984} 	Wald, R.M. \textit{General Relativity}; University of Chicago Press: {Chicago, IL, USA,}
	 1984.
	
	
\bibitem{Parker2009} Parker, L.; Toms, D. \textit{Quantum Field Theory in Curved Spacetime: Quantized Fields and Gravity}; Cambridge University Press: Cambridge, UK, 2009.
	
	
\bibitem{DeWitt1967} DeWitt, B.S. Quantum Theory of Gravity. I. The Canonical Theory. \textit{Phys. Rev.} \textbf{1967}, 160, 1113.
	
	
\bibitem{Hawking1975} Hawking, S.W. Particle Creation by Black Holes. \textit{Commun. Math. Phys.} {\bf 1975}, \textit{43}, 199.
	
	
\bibitem{tHooft1974} 't Hooft, G.; Veltman, M. One-Loop Divergences in the Theory of Gravitation. \textit{Ann. Inst. H. Poincare A} {\bf 1974}, \textit{20}, 69.
	
	
\bibitem{Donoghue1994} Donoghue, J.F. General Relativity as an Effective Field Theory: The Leading Quantum Corrections. \textit{Phys. Rev. D} {\bf 1994}, \textit{50}, 3874.
	
	
\bibitem{Burgess2004} Burgess, C.P. Quantum Gravity in Everyday Life: General Relativity as an Effective Field Theory. \textit{Living Rev. Relativ.} {\bf 2004}, \textit{7}, 5.
	
	
\bibitem{Isham1992} Isham, C.J. \textit{Canonical Quantum Gravity and the Problem of Time. In Integrable Systems, Quantum Groups, and Quantum Field Theories}; Kluwer Academic: Dordrecht, The Netherlands, 1993.
	
	
\bibitem{Rovelli2020} Rovelli, C. Loop Quantum Gravity. \textit{Living Rev. Relativ.} \textbf{2020}, \textit{23}, 5.
	
	
\bibitem{Ashtekar2004} Ashtekar, A.; Lewandowski, J. Background Independent Quantum Gravity: A Status Report. \textit{Class. Quantum Grav.} \textbf{2004}, \textit{21}, R53.
	
	
\bibitem{Kuchar1992} Kuchar, K.V. Time and Interpretations of Quantum Gravity. In \textit{Proceedings of the 4th Canadian Conference on General Relativity and Relativistic Astrophysics}; World Scientific: Singapore, 1992.
	
	
\bibitem{Polchinski1998} Polchinski, J. \textit{String Theory}; Cambridge University Press: Cambridge, UK, 1998.
	
	
\bibitem{Green1987} {Green, M.B.; Schwarz, J.H.; Witten, E. \textit{Superstring Theory}; Cambridge University Press: Cambridge, UK, 1987.} 
	
	
\bibitem{Maldacena1999} Maldacena, J. The Large-N Limit of Superconformal Field Theories and Supergravity. \textit{Int. J. Theor. Phys.} {\bf 1999}, \textit{38}, 1113--1133.
	
	
\bibitem{Susskind2003} Susskind, L. The Anthropic Landscape of String Theory. In \textit{Universe or Multiverse?}; Cambridge University Press: Cambridge, UK, 2007.
	
	
\bibitem{Reuter2012} Reuter, M.; Saueressig, F. Quantum Einstein Gravity. \textit{New J. Phys.} \textbf{2012}, \textit{14}, 055022.
	
	
\bibitem{Eichhorn2019} Eichhorn, A. Status of the Asymptotic Safety Paradigm for Quantum Gravity. \textit{Found. Phys.} \textbf{2019}, \textit{49}, 793.
	
	
\bibitem{Percacci2017} Percacci, R. \textit{An Introduction to Covariant Quantum Gravity and Asymptotic Safety}; World Scientific: Singapore, 2017.
	
	
\bibitem{Sorkin2005} Sorkin, R.D. Causal Sets: Discrete Gravity. In \textit{Lectures on Quantum Gravity}; Gomberoff, A., Marolf, D., Eds.; Springer: New York, NY, USA, 2005; pp. 305--327.
	
	
\bibitem{Loll2019} Loll, R. Quantum Gravity from Causal Dynamical Triangulations: A Review. \textit{Class. Quantum Grav.} \textbf{2019}, 37, 013002.
	
	
\bibitem{Verlinde2017} Verlinde, E. Emergent Gravity and the Dark Universe. \textit{SciPost Phys.} \textbf{2017}, \textit{2}, 016.
	
	
\bibitem{Oriti2014} Oriti, D. The Group Field Theory Approach to Quantum Gravity. \textit{Rep. Prog. Phys.} \textbf{2014}, 77, 066901.
	
	
\bibitem{Krssak2019} Kr\v{s}\v{s}\`ak, M.; van den Hoogen, R.J.; Pereira, J.G.; B\"ohmer, C.G.; Coley, A.A. Teleparallel Theories of Gravity: Illuminating a Fully Invariant Approach. \textit{Class. Quantum Grav.} \textbf{2019}, 36, 183001.
	
	
\bibitem{Bahamonde2023} Bahamonde, S.; B\"ohmer, C.G.; Wright, M.
	Modified teleparallel theories of gravity. \textit{Phys. Rev. D} \textbf{2015}, \textit{92}, 104042.
	
	
\bibitem{Aldrovandi2013} Pereira, J.G.; Aldrovandi, R. \textit{Teleparallel Gravity: An Introduction}; Springer: Dordrecht, The Netherlands, 2013.
	
	
\bibitem{Ferraro2007} Ferraro, R.; Fiorini, F. Modified Teleparallel Gravity: Inflation without Inflaton. \textit{Phys. Rev. D} \textbf{2007}, \textit{75}, 084031.
	
	
\bibitem{33} Bahamonde, S.; Dialektopoulos, K.F.; Levi Said, J. Can Horndeski theory be recast using teleparallel gravity? \textit{Phys. Rev. D} {\bf 2019}, \textit{100}, 064018.
	
	
	
\bibitem{Tseytlin1982} Tseytlin, A.A. Poincar\'e and de Sitter gauge theories of gravity with propagating torsion. \textit{Phys. Rev. D} {\bf 1982}, \textit{26}, 3327.
	
	
\bibitem{LeeNeeman1990} Lee, C.-Y.; Ne'eman, Y. Renormalization of gauge-affine gravity. \textit{Phys. Lett. B} \textbf{1990}, \textit{242}, 59--63.
	
	
\bibitem{Lee1992} Lee, C.-Y. Renormalization of quantum gravity with local $GL(4,\mathbb{R})$ symmetry. \textit{Class. Quant Grav.} \textbf{1992}, \textit{9}, 2001.
	
	
\bibitem{ShapiroTeixeira2014} Shapiro, I.L.; Teixeira, P.M. Quantum Einstein–Cartan theory with the Holst term. \textit{Class. Quant Grav.} \textbf{2014}, \textit{31}, 185002.
	
	
\bibitem{Brandt2024a} Brandt, F.T.; Frenkel, J.; Martins-Filho, S.; McKeon, D.G.C. Quantization of Einstein–Cartan theory in the first order form. {\it Ann. Phys.} \textbf{2024}, \textit{462}, 169607.
	
	
\bibitem{Brandt2024b} Brandt, F.T.; Frenkel, J.; Martins-Filho, S.; McKeon, D.G.C. Renormalization of the Einstein–Cartan theory in first-order form. {\it Ann. Phys.} \textbf{2024}, \textit{470}, 169801.

	
	
\bibitem{Fulling1989} Fulling, S.A. \textit{Aspects of Quantum Field Theory in Curved Space-Time}; Cambridge University Press: Cambridge, UK, 1989.
	
	
\bibitem{Fulling1973} 	Fulling, S.A. Nonuniqueness of Canonical Field Quantization in Riemannian Space-Time.  {\it Phys. Rev. D} {\bf 1973}, \textit{7}, 2850.
	
	
\bibitem{Wald1994} 	Wald, R.M. \textit{Quantum Field Theory in Curved Spacetime}; University of Chicago Press: 
	{Chicago, IL, USA,} 
	 1994.
	
	{
	
\bibitem{Hehl1976} Hehl, F.W.; von der Heyde, P.; Kerlick, G.D.; Nester, J.M. General relativity with spin and torsion: Foundations and prospects. {\it Rev. Mod. Phys.} {\bf 1976}, \textit{48}, 393.
	
	
\bibitem{OgievetskyPolubarinov} Ogievetskii, V.I.; Polubarinov, I.V. Spinors in Gravitation Theory. \textit{Sov. Phys. JETP} {\bf 1965}, \textit{21}, 1093–1100.
	
	
\bibitem{Pitts} Pitts, J.B. Equivalent Theories and Changing Hamiltonian Observables in General Relativity. \textit{Found. Phys.} {\bf 2018}, \textit{48}, 579. 
	}
	
	
\bibitem{AllenJacobson1986} Allen, B.; Jacobson, T. Vector Two-Point Functions in Maximally Symmetric Spaces. \textit{Commun. Math. Phys.} {\bf 1986}, \textit{103}, 669.
		
	
\bibitem{GoroffSagnotti1986} Goroff, M.H.; Sagnotti, A. The Ultraviolet Behavior of Einstein Gravity. \textit{Nucl. Phys. B} {\bf 1986}, \textit{266}, 709.
	
	
\bibitem{Parker1969} Parker, L. Quantized Fields and Particle Creation in Expanding Universes. I. \textit{Phys. Rev.} {\bf 1969}, \textit{183}, 1057.
	
	
\bibitem{Ashtekar1986} 	Ashtekar, A. New Variables for Classical and Quantum Gravity. \textit{Phys. Rev. Lett.} {\bf 1986}, \textit{57}, 2244.
	
	
\bibitem{Rovelli2004} 	Rovelli, C. \textit{Quantum Gravity}; Cambridge University Press: 
	{Cambridge, UK,} 
	 2004.
	
	
\bibitem{Thiemann2007} 	Thiemann, T. \textit{Modern Canonical Quantum General Relativity}; {Cambridge,} 
  2007.
	
	
\bibitem{Perez2013} Perez, A. The Spin Foam Approach to Quantum Gravity. \textit{Living Rev. Relativ.} {\bf 2013}, \textit{16}, 3.
	
	
\bibitem{Ashtekar2006} 	Ashtekar, A.; Pawlowski, T.; Singh, P. Quantum Nature of the Big Bang. \textit{Phys. Rev. Lett.} {\bf 2006}, \textit{96}, 141301.
	
	
	
	
	
\bibitem{DouglasKachru2007} Douglas, M.R.; Kachru, S. Flux Compactification. \textit{Rev. Mod. Phys.} {\bf 2007}, \textit{79}, 733.
	
	{
	
\bibitem{ObukhovPereira2003}  Obukhov, Y.N.; Pereira, J.G. Metric-affine approach to teleparallel gravity. \textit{Phys. Rev. D} \textbf{2003}, \textit{67}, 044016.
	}
	
	
	
	
	
	
	
	
	
	
\bibitem{21} Maluf, J.W. The Teleparallel Equivalent of General Relativity. \textit{Ann. Phys.} {\bf 2013}, \textit{525}, 339--357.
	
	
\bibitem{22} Bengochea, G.R.; Ferraro, R. Dark Torsion as the Cosmic Speed-Up. \textit{Phys. Rev. D} {\bf 2009}, \textit{79}, 124019.
	
	
\bibitem{25} Cai, Y.-F.; Capozziello, S.; De Laurentis, M.; Saridakis, E.N. $f(T)$ Teleparallel Gravity and Cosmology. \textit{Rep. Prog. Phys.} {\bf 2016}, \textit{79}, 106901.
	
	
\bibitem{ColeyLandry2024} Coley, A.A.; Landry, A.; van den Hoogen, R.J.; McNutt, D.D. Spherically Symmetric Teleparallel Geometries. \textit{Eur. Phys. J. C} \textbf{2024},  \textit{84}, 334.
	
	
\bibitem{24} Kr\v{s}\v{s}\'ak, M.; Pereira, J.G. Spin Connection and Local Lorentz Invariance in $f(T)$ Gravity. \textit{Eur. Phys. J. C} {\bf 2015}, \textit{75}, 519.
	
	
\bibitem{Landry2024} Landry, A. Static Spherically Symmetric Perfect Fluid Solutions in Teleparallel $F(T)$ Gravity. \textit{Axioms} {\bf 2024}, \textit{13}, 333.
	
	
\bibitem{Landry2025} Landry, A. Scalar Field Kantowski--Sachs Solutions in Teleparallel $F(T)$ Gravity. \textit{Universe} {\bf 2025}, \textit{11}, 26.
	
	
\bibitem{26} Kavya , N.S.; Mishra, S.S.; Sahoo, P.K.; Venkatesha, V. Can teleparallel $f(T)$ models play a bridge between early and late time Universe, \textit{Monthly Notices of the Royal Astro. Soc.} {\bf 2024}, {\it 532}, 3126. 

	
\bibitem{27} Coley, A.A. \textit{Dynamical Systems and Cosmology}; Kluwer Academic Publishers: Dordrecht, The Netherlands, 2003.
	
	
\bibitem{34} Hayashi, K.; Shirafuji, T. New General Relativity. \textit{Phys. Rev. D} {\bf 1979}, \textit{19}, 3524.
	
	
\bibitem{35} Jim\'enez, J.B.; Heisenberg, L.; Koivisto, T. Teleparallel Palatini Theories. \textit{J. Cosmol. Astropart. Phys.} {\bf 2018}, \textit{8}, 039.
	
	
	
\bibitem{38} Jim\'enez, J.B.; Heisenberg, L.; Koivisto, T. Coincident General Relativity. \textit{Phys. Rev. D} {\bf 2018}, \textit{98}, 044048.
	
	
\bibitem{39} Lazkoz, R.; Lobo, F.S.N.; Ortiz-Banos, M.; Salzano, V. Observational Constraints of $f(Q)$ Gravity. \textit{Phys. Rev. D} {\bf 2019}, \textit{100}, 104027.
	
	
\bibitem{32} Kr\v{s}\v{s}\'ak, M.; Saridakis, E.N. The Covariant Formulation of $f(T)$ Gravity. \textit{Class. Quantum Grav.} {\bf 2016}, \textit{33}, 115009.
	
	
	
	
	
	
\end{thebibliography}
\end{document}